\documentclass[english,aps,showpacs]{revtex4}

\usepackage[dvips]{graphicx}

\begin{document}

\title{Dynamic Density Functional theory for steady currents: \\
Application to colloidal particles in narrow channels}

\author{ F. Penna and P. Tarazona }

\affiliation{Departamento de F\'{\i}sica Te\'orica de la Materia
Condensada,
Universidad Aut\'onoma de Madrid, E-28049 Madrid, Spain \\
}

\begin{abstract}
We present the theoretical analysis of the steady state currents 
and density distributions of particles moving with Langevin dynamics,
under the effects of an external potential displaced at constant rate.
The Dynamic Density Functional (DDF) formalism is used to introduce the 
effects of the molecular interactions, from the equilibrium
Helmholtz free energy density functional. We analyzed the generic
form of the DDF for one-dimensional external potentials and  the limits
of strong and weak potential barriers. The ideal gas case is solved
in a closed form for generic potentials and compared with the
numerical results for hard-rods, with the exact equilibrium free energy.
The results may be of relevance for microfluidic devices, with
colloidal particles moving along narrow channels, if external
driving forces have to compete with the brownian fluctuations
and the interaction forces of the particles.  
\end{abstract}

\maketitle

\section{ Introduction}

Microfluidics, or the manipulation of liquids at the 
pico-liter scale, is an emerging technology with very
interesting perspectives \cite{microflu1,microflu2}.
In particular, the construction of devices which might
control  colloidal fluxes at the level of
single particles, even under the effects of strong 
correlations at high density and in confined geometries
would be of great interest for many applications, and
it is already within the range of experimental works \cite{PRLcha}.
A theoretical description, with quantitative
prediction power, for these systems
would allow the design of devices on a firmer
basis, but such theoretical framework is well established only for 
the properties of systems at thermal equilibrium,
mainly thought the applications of the density functional
formalism (see e.g. \cite{DFF}). 
In this paper we explore the use of a particular extension
of the density functional methods to describe
 some systems out of equilibrium, like fluxes produced by 
the time-dependent external forces in micro-pumps or turn-pikes
which might control single colloidal particles in a flux.
  
The extension of the  density functional formalism (DDF),
from systems at thermodynamic
equilibrium to the dynamics of systems out 
of equilibrium has been a very attractive target for workers
in classical and quantum fluids during the last two decades.
The goal is certainly ambitious and the success has been
limited to some particular types of problems and levels of
description. Thus, the modern
theory of phase transitions dynamics \cite{gendyn} is based on quasi-local
thermodynamic descriptions, which may be considered as 
an empirical extension of the squared gradients density functional
approximation to study the dynamics of a 'coarse grained'
variable. The generic form for such descriptions, for a conserved
'coarse grained' density $\tilde{\rho}({\bf r},t)$ is 
\begin{eqnarray}
\frac{\partial \tilde{\rho} (x,t)}{\partial t}=
\nabla \left[ \Gamma(\tilde{\rho}(x,t)) \ \nabla
\left( \frac{\delta  {\cal F}[\tilde{\rho}]}
{\delta \tilde{\rho}(x,t)} + V(x)  \right) +\xi(x,t) \right],
\label{PTD}
\end{eqnarray}
where ${\cal F}[\rho]$ is the intrinsic the Helmholtz free 
energy functional, approximated by 
\begin{eqnarray}
{\cal F}[\rho] = \int dx \left( f(\rho(x)) + \frac{b}{2} (\nabla
\rho(x))^2 \right), 
\label{SGF}
\end{eqnarray}
$x$ runs over all the space variables, $f(\rho)$ gives the 
free energy per unit volume for bulk systems of homogeneous
density $\rho$, and the coefficient $b$ (which may be taken 
as a function of $\rho(x)$ itself) gives a measure of the 
free energy associated to the inhomogeneity of the density.
$V(x)$ represents any external potential acting on the 
particles.

   The density functional approximation (\ref{SGF}) has
been extensively used to describe equilibrium systems
with smooth spatial inhomogeneities, like the liquid-vapor
interfaces \cite{Evans1,Rowl}, although it cannot describe systems
with stronger modulations of the local density, like the 
layered structures of fluids near walls \cite{Evans2}.
The use of this ${ \cal F}[\rho]$ for dynamics requires
to specify two other empirical elements, the mobility
$\Gamma(\tilde{\rho})$ which relates the density current
to the local gradient of the thermodynamic chemical 
potential, and the random noise term $\xi(x,t)$ 
usually taken as a gaussian (uncorrelated) stochastic field
which allows the systems go over thermodynamic barriers,
e.g. in the process of nucleation of liquid drops from 
oversaturated vapors. Both the mobility as a function of
the coarse grained density, and the amplitude of the
gaussian noise have to be taken as empirical mesoscopic
parameters, which control the relaxation process of the
coarse grained density, but without a direct connection to
the molecular level. Therefore, and although equation (\ref{PTD})
has been very useful to analyze different aspects of the
phase transition dynamics \cite{gendyn}, it cannot be used
down to a molecular level of description, nor can it be 
extended to make use of the very accurate equilibrium 
density functional approximations \cite{Evans2}
which go much beyond the simple functional form (\ref{SGF})
in the description of equilibrium systems. Other
approaches to define and use a DDF theory have used a
more microscopic descriptions of the density distribution
\cite{Kawa1, Dean, Kawa2} but still keeping some coarse graining
and a random noise term to overcome the energetic barriers
of the coarse grained free-energy density functional. 

A different DDF approach, which a clear
connection between the molecular hamiltonian and a 
dynamic density functional equation
may be obtained \cite{OurJCP,OurJPCM} from the hypothesis
of purely relaxative Langevin dynamics at molecular level;
i.e. if the positions of the particles $r_i(t)$
(for $i=1,2,...,N$) evolve in time according to the equation
\begin{eqnarray}
\frac{d r_i(t)}{d t} = - \Gamma_o \  \nabla_{i}
{\cal U}(r_1,...,r_N) + \ \xi_{i} , 
\label{Langevin}
\end{eqnarray}
where ${\cal U}(r_1,...,r_N)$ is the potential energy 
which includes the interactions between the $N$ particles
in the system, and also any external potential acting on them.
The molecular mobility $\Gamma_o$ gives the (assumed constant) ratio
between the mean velocity of the particles under a constant external
force. When the theory is applied to colloidal particles 
moving in a fluid, the constant $\Gamma_o$ may be obtained by Stokes
formula in terms of their radius and the viscosity of the fluid
matrix. The second term in the bracket in (\ref{Langevin}),
$\xi_{i}$ represents a random force acting on particle $i$,
with may be interpreted as the zero-average stochastic effect
of the fluid matrix molecules on the colloidal particles.
This is taken as gaussian uncorrelated function for each
particle, but contrary to the random noise
$\xi(x,t)$ in (\ref{PTD}), its amplitude is perfectly defined
by Einstein relation, as twice the diffusion constant 
$D= \Gamma_o k_{\hbox{\tiny B}} T$.
The natural reduced units for the problem are 
$k_{\hbox{\tiny B}} T \equiv 1/\beta$
for the energy, molecular diameter $\sigma$ for distance,
$\beta \sigma^2/\Gamma_o$ for time, and hence  
$\Gamma_o \ k_{\hbox{\tiny B}} T / \sigma$ for the velocity. 

 Using the standard tools of stochastic analysis \cite{Gardiner}
and the equilibrium density functional formalism is possible 
to transform (\ref{Langevin}) into a Dynamic Density Functional 
(DDF) equation, 
\begin{eqnarray}
\frac{\partial \tilde{\rho} (x,t)}{\partial t}= \Gamma_o
\ \nabla \left[ \tilde{\rho}(x,t) \ \nabla
\left( \frac{\delta{ \cal F}[\tilde{\rho}]}
{\delta \tilde{\rho}(x,t)} + V(x,t) \right) \right],
\label{DDF}
\end{eqnarray}
within the following assumptions:

i) The density  $\tilde{\rho}(x,t)$, following the 
dynamics of the DDF equation (\ref{DDF}), has to be interpreted 
as the average of the instantaneous density 
$
\hat{\rho}(x,t)= \sum_{i=1}^N \delta(x-r_i(t)),
$
over an statistical ensemble of realizations for the
random noise $\xi(t)$ for the integration of the Langevin
equation, from given initial conditions 
$\tilde{\rho}(x,0)=\tilde{\rho}_{\hbox{\tiny inic}}(x)$. The
temperature $T$ is assumed constant and fixed by the bath.

ii) The correlations between the colloidal particles 
in the dynamical system, at any instant $t$, 
are assumed to be identical to
those in the equilibrium systems, which would be obtained
by applied the appropriate external field to the system 
so that its equilibrium density distribution, $\rho_{\hbox{\tiny eq}}(x)$
would be given precisely by the instantaneous value
of  $\tilde{\rho}(x,t)$. These equilibrium correlations
are those given by the density functional giving the intrinsic
equilibrium Helmholtz free energy of the fluid, ${ \cal F}[\rho]$.
This equilibrium free energy functional, for classical fluids, 
has an explicit dependence from the ideal gas entropy contribution,
$$
{\cal F}_{id}[\rho]=k_b T \int dx  \rho(x) \left( log(\rho(x)) - 1 \right)
$$
and a contribution, 
$\Delta {\cal F}[\rho] \equiv {\cal F}[\rho] - {\cal F}_{id}[\rho]$,
from the particle interactions, which is known exactly
only for a few systems; but approximations with increasing accuracy
have been developed for others and give excellent results for the
equilibrium properties in many cases. 

   Beside having a perfectly defined mobility factor,
as the density $\tilde{\rho}(x,t)$ times  the molecular
mobility in the Langevin equation, instead of the 
generic function $\Gamma(\tilde{\rho}(x,t))$ in (\ref{PTD});
the main difference between that semi-empirical equation 
and the DDF equation (\ref{DDF}) is the lack of any random noise
term in later, which becomes a deterministic functional
equation instead of the usual stochastic functional equation
(\ref{PTD}). The averaging over the random force term in the 
original Langevin equation (\ref{Langevin}), required to get
$\tilde{\rho}(x,t)$ with the interpretation $i)$ above,
gives the classical ideal gas contribution, ${ \cal F}_{id}[\rho]$ to 
${\cal F}[\rho]$ in the functional derivative of
(\ref{DDF}), with the exact prefactor $k_b T$ from the precise
amplitude of the random force $\xi_i(t)$; and any extra random
noise term in  (\ref{DDF}) would produce the overestimation 
of the ideal gas entropy. The interactions between the 
particles, included in the derivative
of ${ \cal U}$ in (\ref{Langevin}) give the 
contribution $\Delta { \cal F}$, to the intrinsic free 
energy functional ${ \cal F}[\rho]$ in (\ref{DDF}),
while the contribution to the potential energy from external
fields appears explicitly as $V(x,t)$.
We may expect that the accuracy of the approach to include the
effects of the interactions would improve with the use of
accurate forms of ${ \cal F}[\rho]$, which include the non-local
functional dependence in a much better way than (\ref{SGF}).

 In particular, we may think of colloidal particles,
interacting as hard-core molecules of diameter $\sigma$
and confined along a linear  channel, as already explored
experimentally \cite{PRLcha}, which may be theoretically
described as hard-rods in one dimension, for which the
exact equilibrium free energy density functional is known
\cite{Percus}. The solution of (\ref{DDF}) for hard-rods
has already been studied for the relaxative dynamics
under several static external potentials \cite{OurJCP}.
An important conclusion of that work was that the use of 
the exact  ${ \cal F}[\rho]$ made unnecessary the 
inclusion of the random noise term in the DDF equation,
since the exact free energy is always a
convex functional, with only one minimum. The existence
of activation barriers, between two different minima of 
the free energy, which should be overcome by the stochastic
noise is an artifact of using an approximated form 
for ${\cal F}[\rho]$. Thus, very long relaxation times which
in the solution of (\ref{PTD}) would be interpreted
as typical of systems with a high free energy barrier
between two minima, are recovered from the deterministic
equation (\ref{DDF}) without any free energy barrier,
just from the particle correlations described through
the exact free energy ${\cal F}[\rho]$. In this context,
if we take colloidal particles of size $\sigma \approx 10^{-7} m$,
with mobility $\Gamma_o$ given by Stokes law in water, at 
room temperature, the natural unit for velocity is
$\Gamma_o k_{\hbox{\tiny B}} T/\sigma \approx 10^{-4} m/s$, in 
a reasonable range for potential barrier shifts in 
microfluidic devices.
 
   In this paper we analyze the results of (\ref{DDF}) for
a different class of problems, hard-rods particles in 
time-dependent external potentials $V(x,t)$. In particular
we start with systems in which the external potential
moves with constant rate $c$ with respect to the static
framework in which equation (\ref{Langevin}) applies.
 These problems would apply to colloidal particles 
moving along narrow channel under the force created
by a moving electrical potential or optical twist;
which may be used as a starting point for the design 
of more complex devices, like micropumps or turn-pikes.
The plan of the paper is the following, in the next section
we analyze the generic structure of the solutions
of the DDF equation (\ref{DDF}) for the stationary states
with density $\tilde{\rho}(x,t)=\tilde{\rho}(x-ct)$
moving along the external potential $V(x,t)=V(x-ct)$.
In section 3 we get an explicit solution for the ideal
gas case, when the interaction term  ${\cal F}[\rho]$ may
neglected; this solution helps to give a qualitative 
classification of the possible behaviour, in terms 
of the barriers of the external potential and the
rate of displacement. In section 4 we explore, numerically
the solutions of the DDF when the packing effects are
important and described by the exact density functional 
at equilibrium for hard rods.  We end with a general discussion 
on the observed qualitative behaviour and generic trend
in the dependence with the physical parameters: the height
and shape of the external potential barriers, the shift rate 
of these barriers with respect to the static framework for
the relaxative dynamics and the effects of molecular packing.

\section{Stationary solutions of the DDF}

The generic DDF equation (\ref{DDF}) may be interpreted
as the exact continuity equation between the density
of particles, $\tilde{\rho}(x,t)$ and their current $j(x,t)$,
\begin{eqnarray}
\frac{\partial \tilde{\rho} (x,t)}{\partial t}=
- \ \nabla j(x,t),
\label{cont}
\end{eqnarray}
together with the approximated DDF closure for the 
current,
\begin{eqnarray}
j(x,t)= - \ \Gamma_o
\  \tilde{\rho}(x,t) \ \nabla
\left( \frac{\delta{ \cal F}[\tilde{\rho}]}
{\delta \tilde{\rho}(x,t)} + V(x,t) \right),
\label{jota}
\end{eqnarray}
in terms of the equilibrium free energy density functional.

  In the presence of a shifting external potential
$V(x,t)=V(x-ct)$ we look for stationary solutions 
of the form $\rho(x,t)=\rho(x-ct)$, and 
$j(x,t)=j(x-ct)$, so that the continuity equation (\ref{cont}) 
implies that $c \nabla \rho(x,t)= \nabla j(x,t)$, which for any 
problem in one dimension (1D) gives
\begin{eqnarray}
j(x,t)= c \ \rho(x,t) - \Delta, 
\label{jota1}
\end{eqnarray}
with an arbitrary integration constant $\Delta$, which
gives the difference between the actual current and that
which would be produce by the rigid shift of the 
density profile at the same velocity $c$ as the potential.
This counter-drift term $\Delta$ cannot be determined
from the DDF equation (\ref{DDF}), and it would be obtained
only through the boundary conditions that we apply to the 
system.     

 Using now equations (\ref{jota}) and (\ref{jota1}) we
may obtain a closed functional equation for the stationary
density $\tilde{\rho}(x,t)\equiv \tilde{\rho}(\hat{x})$,
where we defined the variable $\hat{x}=x-ct$, moving
with the external potential  $V(x,t)\equiv V(\hat{x})$;
\begin{eqnarray}
 \tilde{\rho}(\hat{x}) \ \nabla
\left( \frac{\delta{ \cal F}[\tilde{\rho}]}
{\delta \tilde{\rho}(\hat{x})} + V(\hat{x}) \right) +
  \frac{c}{\Gamma_o} \ \tilde{\rho}(\hat{x})
= \frac{ \Delta}{\Gamma_o}.
\label{DDFeq}
\end{eqnarray}
 
In 1D problems this expression may be formally integrated to give
\begin{eqnarray}
 \frac{\delta{ \cal F}[\tilde{\rho}]}
{\delta \tilde{\rho}(\hat{x})} + V(\hat{x}) +
  \frac{c}{\Gamma_o} \ (\hat{x}-\hat{x}_o)  -
\frac{ \Delta}{\Gamma_o} \int_{\hat{x}_o}^{\hat{x}} \frac {d\hat{x}'}
{\tilde{\rho}(\hat{x}')}=\mu,
\label{DDFeq1}
\end{eqnarray}
where $\mu$ is the integration constant, linked to the arbitrary
value of $\hat{x}_o$ in the lower limit of the integral.

  In the case of static external potentials, $c=0$, the 
stationary density becomes the equilibrium density, and
current $j$ has to vanish everywhere, so that the counter-drift constant 
$\Delta$ in (\ref{jota1}) vanishes, and
(\ref{DDFeq1}) becomes the usual Euler-Lagrange equation
for the equilibrium density, $\rho(x)=\tilde{\rho}(\hat{x})$,
with the integration constant $\mu$ as the equilibrium
chemical potential. Thus, in the dynamic case
($c \neq 0$) the integration constant $\mu$ in (\ref{DDFeq1})
will still be used to control the total number of particles
in the system, since  for a given external potential $V(\hat{x})$ and drifting
rate $c$, that functional equation should have different solutions
representing dynamic systems with different mean densities.
The presence of a second integration constant $\Delta$ in
(\ref{DDFeq1}) is characteristics of dynamic systems, as it
has to vanish in the limit $c=0$. To determine the value of
$\Delta$ we may consider a further symmetry assuming that the
drifting potential $V(\hat{x})$ and hence the stationary 
density $\tilde{\rho}(\hat{x})$ are periodic functions 
with spacial period $L$ (which imply a time period $T=L/c$).
In that case, comparing the values of (\ref{DDFeq1}) for
$\hat{x}$ and $\hat{x}+L$ we get
\begin{eqnarray}
  \frac{c}{\Gamma_o} \  L =
\frac{ \Delta}{\Gamma_o} \int_{\hat{x}}^{L+\hat{x}} \frac {d\hat{x}'}
{\tilde{\rho}(\hat{x}')},
\label{periodic}
\end{eqnarray}
for any value of $\hat{x}$, which implies 
\begin{eqnarray}
\Delta= \frac {c}{\langle\tilde{\rho}^{-1}\rangle},
\label{Delta}
\end{eqnarray}
and the stationary current is
\begin{eqnarray}
j(\hat{x})= c \ \left(  \tilde{\rho}(\hat{x})- 
\frac {1}{\langle\tilde{\rho}^{-1}\rangle}
\right),
\end{eqnarray}
where the brackets represent the spacial average over a full period
of the inverse local density. The last expression makes clear
that for static periodic potential, $c=0$,
the only stationary solution of (\ref{DDF}) is the equilibrium
state, with the current vanishing everywhere.
 
The stationary density distributions and currents 
created by any periodic external potentials
moving with a constant rate $c$ may now be obtained from 
the solutions of (\ref{DDFeq1}) and (\ref{Delta}), with
the remaining integration constant $\mu$ playing the
same role as in the Euler-Lagrange equation, to control
the mean density of the system. Before including an 
explicit form for the interaction between the particles,
i.e. for the term $\Delta {\cal F}[\rho]$ in free
energy density functional, we may still get some generic
analysis for particular limits. The case of a shifting external
potential limited to a finite region of an infinite 1D
system may be obtained as the $L \rightarrow \infty$ limit
of the periodic case. In that case, unless $V(\hat{x})$ includes
infinite barriers where $\rho(\hat{x})=0$ (as explored in section 2a below)
we get $\Delta= c  \rho_o$, determined by the density $\rho_o(\mu)$ of the 
uniform system at chemical potential $\mu$, and equation 
(\ref{DDFeq1}) may be written as:
\begin{eqnarray}
 \frac{\delta{ \cal F}[\tilde{\rho}]}
{\delta \tilde{\rho}(\hat{x})} + V(\hat{x}) +
  \frac{c}{\Gamma_o}
\int_{-\infty}^{\hat{x}}d\hat{x}' \left( 1 - \frac { \rho_o}
{\tilde{\rho}(\hat{x}')} \right) =\mu.
\label{DDFeq5}
\end{eqnarray}
The current vanishes away from the region crossed by the external 
potential, and the integrated  flow along the 
system becomes 
\begin{eqnarray}
\int_{-\infty}^{\infty}d\hat{x} j(\hat{x}) = 
c \int_{-\infty}^{\infty} d\hat{x} (\tilde{\rho}(\hat{x}) -\rho_o)=
c (N-N_o),
\end{eqnarray}
i.e. c times the excess of particles generated by the shifting external
potential, from the solution of (\ref{DDFeq5}). This expression is
just an integrated form of the continuity  equation since the 
total transport of molecules is precisely the
steady shift of the stationary density perturbation $\tilde{\rho}(\hat{x})-
\rho_o$.

\subsection{  Complete drift limit}

 Let us consider a moving external potential which includes
a periodic series of infinite barriers, where the density 
$\tilde{\rho}(\hat{x})$
which solves (\ref{DDFeq}) has to become zero. Then the average
of the inverse density becomes infinite, whatever values 
$\tilde{\rho}(\hat{x})$ takes outside the barrier. The
counter-drift term $\Delta$ vanishes and the current becomes
equal to the density times the velocity of the external potential
$c$. This is a regime of complete drift, where all the fluid particles
trapped between two consecutive barriers of the external potential
have to follow the movement of these barriers, so that 
$j(\hat{x})=c \ \tilde{\rho}(\hat{x})$. Moreover, the
functional equation (\ref{DDFeq1}) for the density distribution 
in that regime becomes equivalent to the equilibrium 
Euler-Lagrange equation
\begin{eqnarray}
 \frac{\delta{ \cal F}[\tilde{\rho}]}
{\delta \tilde{\rho}(\hat{x})} + V_k(\hat{x}) =\mu,
\label{ELc}
\end{eqnarray}
with a 'kinetic' external potential 
$V_k(\hat{x})=V(\hat{x})+ c/\Gamma_o \ \hat{x}$,
i.e. with the true external potential plus a linear term proportional
to $c$. In order to simplify the notation, we use a reduced
drift rate with inverse length units, as $\bar{c}=\beta 
c/\Gamma_o$.

 The stationary density profile follows a 
variational principle equivalent to that of the equilibrium 
density functional formalism; i.e. $\tilde{\rho}(x,t)$, described in terms
of the drifted variable $\hat{x}=x-ct$, is such that it minimizes
the total 'kinetic' grand potential energy of the system
$$
\Omega[\rho]={\cal F}[\rho] + \int dx \rho(x) (V_k(x)-\mu).
$$
The constant slope $c/\Gamma_o$ of the kinetic contribution to $V_k(x)$
includes the extra force to keep the particles moving with
mean velocity $c$, according with their Langevin dynamics (\ref{Langevin}).
The main advantage of this limit is that the solution of the DDF equation
may be obtained with the well developed methods used for the equilibrium
DFF, like conjugated gradients minimization of $\Omega[\rho]$.

 When the periodic shifting potential $V(\hat{x})$ has high 
(but not infinite) barriers, we may expect that the stationary
profiles $\tilde{\rho}(\hat{x})$ would still be very close
to the solutions of the equivalent equilibrium equation (\ref{ELc}),
since everywhere outside the barriers  
$\tilde{\rho}(\hat{x}) \gg \langle \tilde{\rho}^{-1}\rangle^{-1}$ and the
contribution from the integral in  (\ref{DDFeq1}) would be
nearly constant (and hence transferable to $\mu$). However,
the contribution of that integral over a full period has to
exactly cancel the contribution of the linear term 
$\bar{c} \ \tilde{x}$, to get a periodic equation, which
is achieved by the integral giving a descending staircase
shape, with smoothed steps at  the barriers, and 
nearly planar plateau between them. Thus, with the appropriate 
choices of $\mu$ in each case, the solutions
of (\ref{ELc}) over a single period,  would be similar to those
of the DDF equation (\ref{DDFeq1}) everywhere except in the
low density region at the high potential barriers, where the
true stationary profile has to recover the periodicity.
Only in the strict limit of infinite barriers
the equivalent equation  (\ref{ELc})
may be regarded as exact, when used for the interval between
point at consecutive barriers where $\tilde{\rho}(\hat{x})=0$,
as the role the integral is only to provide a constant shift
of the chemical potential between each period.

 Moreover, even if the  potential barriers of $V(\hat{x})$ 
are apparently very high (on the $k_bT$ scale) the 
quasi-complete drift regime 
described above may be frustrated when the drift rate $c$ becomes
too high, since the energy difference between the maximum and the
minimum of $V_k(\hat{x})$ diminishes with $c$, the 
effective barriers of the kinetic potential may become comparable
with  $k_bT$, and the counter-drift current grow. Thus, we
expect a low velocity regime, in which the velocity of
the particles $v=j(\hat{x})/\tilde{\rho}(\hat{x})$ is 
exactly $c$, and a saturation range in which further increase
of $c$ does not produce the increase of $v$.

\subsection{Weak external potentials}

  We consider now the opposite limit when the external potential
$V(\hat{x})$ is very weak everywhere, so that it may be treated as
a small perturbation over an otherwise uniform fluid.
We may then expand the density around its mean value,
$\tilde{\rho}(\hat{x})=\rho_o + \delta \tilde{\rho}(\hat{x})$,
with $\rho_o = \langle \tilde{\rho}(\hat{x})\rangle$, so that
$\langle 1/\tilde{\rho}(\hat{x})\rangle=\rho_o^{-1}- 
\langle \delta \tilde{\rho}(\hat{x})^{2}\rangle/\rho_o^{3}+...$; 
the current becomes now
$$
j(\hat{x})=c \left( \delta \tilde{\rho}(\hat{x})+ 
\frac {\langle \delta \tilde{\rho}(\hat{x})^{2}\rangle}{\rho_o}+
{\cal O}^{3}(\delta \tilde{\rho}) \right),
$$
and its average is
$\langle j \rangle= c \ \langle \delta \tilde{\rho}(\hat{x})^{2}\rangle/\rho_o+{\cal O}^{3}(\delta \tilde{\rho})$. Thus, a shifting weak external 
potential $V(x-ct)=V(\hat{x})$ cannot produce the drift of the 
bulk fluid, the local current is proportional to $c$ times the
perturbation over the homogeneous fluid, with regions of 
positive and negative sign, and its average goes only as the
density perturbation squared; i.e. the propagating waves in the
density, $\tilde{\rho}(x-ct)= \tilde{\rho}(\hat{x})$, produce 
a very small mean propagation of particles.

So far in this subsection, we have only used the continuity
equation (\ref{cont}) and the requirement of having a
periodic stationary density (\ref{periodic}) of the DDF equation,
to get the current in terms of the density perturbation. 
To get a relationship between 
the weak external potential, $V(\hat{x})$, and the 
density perturbation $\delta \tilde{\rho}(\hat{x})$
we may use the expansion of the DDF equation (\ref{DDFeq1})
around $\tilde{\rho}(\hat{x})=\rho_o$, as a generalization
of the equilibrium linear response theory \cite{Evans1} to the
dynamic case. In terms of the Fourier transform of
the density perturbation $\delta \tilde{\rho}(q)$
and the external potential $V(q)$, the linear term
of (\ref{DDFeq1}) gives,
\begin{eqnarray}
\delta \tilde{\rho}(q) = - \rho_o \   \frac{ \beta V(q)}
{ [S(q,\rho_o)]^{-1}+ i \ \hat{c}/q},
\label{linear}
\end{eqnarray}
where $S(q,\rho_o)$ is the equilibrium structure factor 
of the homogeneous system, which  is proportional to the 
response function of the system to external field,
as given by the second functional derivative of ${ \cal F}[\rho]$,
by the standard DF analysis \cite{Evans1,Hansen}. 
In the long wavelength limit it is given by the
isothermal compressibility , 
$S(0,\rho_o)= k T \rho_o \chi_{\hbox{\tiny T}}(\rho_o)$
while $S(q,\rho_o)= 1$  in the short wave limit, 
$q \rightarrow \infty$.
The second term in the denominator of
(\ref{linear}) comes from the drift and counter-drift terms in
(\ref{DDFeq1}), which give and imaginary part to the kinetic
response function.

If we consider a weak external potential with the form of
an harmonic propagating wave, $V(x-c \ t)= V_q \  cos(q \ \hat{x})$,
the density would be 
$ \tilde{\rho}(x-c \ t)= \rho_o-\rho_q cos(q \ \hat{x}-\theta)$,
with amplitude and phase shift,
$$
\rho_q= - \rho_o \frac{\beta V_q}
{\left[  [S(q,\rho_o)]^{-2} +
 \left( \bar{c}/q \right)^{2} \right]^{1/2} } \ \ \ \hbox{and} \ \ \ 
\theta=tan^{-1}\left(\frac{\bar{c} \  S(q,\rho_o)}{ q} \right).
$$
For low drift rates the
density profile will be similar to that created by a
static external potential, with the maximum density at the
minima of $V(\hat{x})$ and vice versa. When the external potential
moves fast enough the effect of the imaginary part in the 
denominator of (\ref{linear}) puts the density perturbation
out of phase with the external potential, and the maximum of
$\tilde{\rho}(\hat{x})$ moves toward the region with 
maximum external force, with $\theta \rightarrow \pi/2$,
but the amplitude decreases as $\rho_q \sim 1/c$ for large $c$.

  The mean velocity of the particles, in same reduced units as
$\bar{c}$, for any weak periodic potential may be written as a 
Fourier sum with the quadratic contributions of each Fourier component
of the potential: 
\begin{eqnarray}
\bar{v}= \frac{\beta \langle j \rangle}{\Gamma_o \rho_o}=
 \ \bar{c} \ \sum_q \frac{ (\beta V(q))^2} 
{[S(q,\rho_o)]^{-2} + \left( \bar{c}/q \right)^{2}}.
\label{vweak}
\end{eqnarray}

The contribution of the  short wavelength components, with 
$\bar{c} \ S(q,\rho_o) \ll q$, is similar to the 
quasi-static limit ($\theta \approx 0$),
with $\bar{v}$ linear in 
$c$ and quadratic with the static response $ S(q,\rho_o) \ V(q)$;
in the opposite limit, $\bar{c} \ S(q,\rho_o) \gg q$, the 
kinetic contribution dominates ($\theta \approx \pi/2$) and
the contribution to $\bar{v}$ is proportional to $(V(q) q)^2/c$, i.e.
decays with $c$, is independently of static structure factor
and goes proportional to the mean squared
force $\langle ( V'(x) )^2 \rangle$, rather than to the 
mean squared potential. For each Fourier component of $V(x)$, the 
threshold between the linear growth,
$\bar{v} \sim \bar{c}$, for low $\bar{c}$ and the decay, 
$\bar{v} \sim 1/\bar{c}$, for
large $\bar{c}$, would give the optimum shift rate 
$\bar{c}_{\hbox{\tiny max }} \sim q/ S(q,\rho_o)$ to get the maximum
mean velocity. In the ideal gas limit, $S(q,\rho_o)=1$,
this optimum value is $\bar{c}=q$, becoming very small in the 
long wavelength limit. For correlated systems, there would 
be a non monotonic dependence of the maximum $\bar{v}$ with
$c$, depending on $q$ due to the static structure factor.

\section{ Solution of the DDF equation for the ideal gas}

In the previous section we have obtained some generic 
limiting behaviour for the solutions of the DDF equation.
We turn now to explore the general solution of (\ref{DDFeq1}),
first in absence of particle interactions, and in the following
section including the effects of molecular packing as 
one dimension  hard rods.  

  The ideal gas case corresponds to take $\Delta {\cal F}[\rho]=0$
and it should be the exact low density limit of any fluid system.
Moreover, the assumptions to get (\ref{DDF}) from (\ref{Langevin})
become irrelevant, and the DDF equation may be consider as an 
exact Fokker-Planck equation for the density distribution. 
The equation (\ref{DDFeq})  becomes
\begin{eqnarray}
 \ \nabla  \tilde{\rho}(\hat{x}) +
 \tilde{\rho}(\hat{x})
\left(\beta \nabla V(\hat{x})+ \bar{c}  \right)
= \frac{\beta \Delta}{\Gamma_o}. 
\label{DDFid}
\end{eqnarray}
 When $\Delta$ is zero (i.e. in the static limit, $c=0$, or in
the complete drift limit described above), the homogeneous 
linear differential equation is directly solved to give
$ \tilde{\rho}(\hat{x})= A \ exp(-\beta V_k(\hat{x}))$
in terms of the Boltzmann factor of the kinetic potential
$ \beta V_k(\tilde{x})= \beta V(\tilde{x})+\bar{c} \tilde{x}$.
The arbitrary constant $A$ is the activity, playing the
same role as the chemical potential $\mu$ in (\ref{DDFeq1})
to control the total number of particles.

 The general solution of the inhomogeneous equation with
$\Delta \neq 0$ is obtained, by the usual Green function method. 
For periodic external potentials, with period $L$, it has 
the generic form,
\begin{eqnarray}
 \tilde{\rho}(\hat{x}) = A \  e^{-\beta V_k(\hat{x})} \left(1 +
\frac{ e^{\bar{c} L} -1 }
{\int_0^{L} dy \ e^{\beta V_k(y)}}
\ \int_0^{\hat{x}} dy \ e^{\beta V_k(y)}
\right),
\label{idsol}
\end{eqnarray}
which may be evaluated in terms of single integrals of the 
inverse Boltzmann factor of $V_k(x)$.  
The counter-drift term $\Delta$, is $\Gamma_o A/\beta$ times
the prefactor of the integral, inside the bracket.

  The average velocity of the particles, defined as the 
ratio between the mean current $\langle j(\hat{x})\rangle$ 
and the mean density $\langle \rho(\hat{x})\rangle$, and in 
the same reduced units as $\bar{c}$ is
\begin{eqnarray}
\bar{v} \equiv \frac{ \beta \langle j(\hat{x})\rangle}
{\Gamma_o \langle \rho(\hat{x})\rangle}=
\bar{c}  \  \frac{\xi(\bar{c})}{1+\xi(\bar{c})},
\label{vide}
\end{eqnarray}
with the dimensionless  auxiliary function 
\begin{eqnarray}
\xi(\bar{c})= \frac{\bar{c}}{L} \int_0^L dx \ e^{-\bar{c} x}
\int_0^x dy \left( e^{\beta(V(y)-V(x))} - 1 \right) e^{\bar{c} y} 
\nonumber
\\ 
+ \ \frac{\bar{c}}{L} \frac{e^{- \bar{c} L}}{1-e^{- \bar{c} L}}
\left( \int_0^L dx \ e^{-\beta V(x) -\bar{c} x} \right)
\left( \int_0^L dy \  e^{\beta V(y) +\bar{c} y} \right) \ 
-\frac{1-e^{- \bar{c} L}}{\bar{c} L}.
\label{xide}
\end{eqnarray}

  It is instructive to analyze the structure of this equation for
potential barrier $V(x)$ which is zero
outside an interval, taken as  $0 \leq x \leq a$, much
shorter than the periodic length of the system $L$.
When the maximum height of the potential barrier, $\beta V(x)$,
is well above the maximum kinetic term $\bar{c} L$, the
results of (\ref{idsol},\ref{vide}) is always within the 
quasi-complete drift regime described above, with
$\Delta \ll c \ \langle \rho \rangle$ and $\bar{v} \approx \bar{c}$.
For larger drifting rates, when $\bar{c} L$ becomes much 
larger than the maximum of $\beta V(x)$, the main
contribution to $\xi(\bar{c})$  comes always from the first 
term in the right hand side of (\ref{xide}), with a delicate
balance between the fast growing exponentials in the integrals. 

The simplest case of a square barrier with $V(x)=V_o$ in an interval 
of width $a$ and $V(x)=0$ outside, may be solved analytically 
to give $\xi(\bar{c})$ and $\bar{v}$. For large drifting rates,
$\bar{c} \ min[a,(L-a)] \gg 1$, we get
\begin{eqnarray}
\bar{v} \approx \bar{c} \ \frac{e^{\beta V_o} -1}{\bar{c} L + e^{\beta V_o} -1}.
\label{videsq}
\end{eqnarray}
 Therefore, a square potential barrier of height $V_o$
would be enough to produce the complete shift of the ideal particles
up to  (reduced) drifting rates $\bar{c}$ well below  $(e^{\beta V_o} -1)/L$.
For  $\bar{c}$  beyond this limit 
the mean velocity would saturate  to a constant value
$\bar{v} \approx e^{\beta V_o}/L$ for $\beta V_o \gg 1 $
and $\bar{v} \approx \beta V_o/L$ for $\beta V_o \ll 1 $.  
In Figure 1 we compare the values of  $\bar{v}$ for square barriers,
parabolic barriers, $V(\hat{x})= V_o (1-(2 \hat{x}/w)^2)$
for $-w/2 \leq  \hat{x} \leq w/2$, and  gaussian  barriers
$V(\hat{x})= V_o exp[- \hat{x}^2/(2 \alpha^2)]$. At low $\bar{c}$
the three barriers give similar results, with $\bar{v}$ proportional
to $\bar{c}$ both for high barriers (Fig. 1(a)) and low barriers
(Fig. 1(b)). The fortes case corresponds to the complete
drift regime  $\bar{v}=\bar{c}$, predicted in section 2(a) for
large enough values of $\beta V_o$; while the second corresponds to
the linear response result $\bar{v} \sim \bar{c} \ (\beta V_o)^2/L$
 predicted in section 2(b) for low $\bar{c}$. At higher values of 
the shifting rate
$\bar{c}$, the results of the square barrier differ qualitatively  
from those of the parabolic and gaussian barriers. The two later
reach a maximum value of the mean velocity, $\bar{v}_{max}=
\bar{v}(\bar{c}_{max})$, and then  decay as $\bar{v} \sim 1/\bar{c}$, 
while the square barrier produce a monotonous increase of
$\bar{v}$ up to a constant value for large $\bar{c}$.
The comparison of the
results with different barrier shapes and parameters shows that
the most relevant aspect of $V(\hat{x})$ is the maximum slope,
rather than the barrier width or the barrier height. Thus,
reducing the width of a parabolic barrier, while keeping unchanged
its maximum value $V_o$, increases
rather than decreases the maximum value of the mean velocity,
since it increases the maximum force acting on the particles. 
In this sense the square barrier model should be consider
unphysical, since the infinite force acting at the barrier borders
creates the saturation of $\bar{v}$ at large $\bar{c}$. Choosing
the parameters of the parabolic and gaussian barriers to have
similar maximum slopes, produces very similar results for 
$\bar{v}(\bar{c})$, as shown in Figure 1 for the 
empirical choice of $w=5.5 \alpha$ and the same barriers height, 
$V_o$.  This robustness of the results to the particular form 
of $V(x)$, as far as it does nor include unrealistic discontinuities,
might help to eventual comparisons of the theory with experimental 
results.

 The weak  potential limit (\ref{vweak}), integrated over the
Fourier components of the different barriers $V(\hat{x})$,
may be compared with the exact results obtained from 
(\ref{vide},\ref{xide}). The qualitative difference 
between the sharp square barrier and the gaussian or parabolic 
models is a consequence of the slow, $V(q) \sim 1/q$,
decay of the external potential Fourier components for the 
discontinuous square barrier; while the first derivative discontinuity
(i.e. $V(q) \sim 1/q^2$) for the parabolic barrier is already weak enough
not to change the qualitative dependence of $\bar{v}(\bar{c})$ with respect
to the smooth gaussian barrier.  For the later, and assuming that
the distance $L$ between barriers is large compared with the barrier width 
$\alpha$, we may perform the sum over Fourier components in  (\ref{vweak})
as an integral to get an analytic form for the  mean velocity within 
the linear response approximation:
$$
\bar{v}(\bar{c})= \frac{(\beta V_o)^2}{L} \alpha \bar{c} \big( 
\sqrt{\pi} - \pi  \alpha \bar{c} \  e^{ \alpha \bar{c}} \ \  
Erfc[ \alpha \bar{c}] \big),
$$
in terms of the complementary error function which leads to
$$\bar{v}(\bar{c}) =  
\frac{\sqrt{\pi} \ (\beta V_o)^2}{L} \left(\frac{1}{2 \alpha \bar{c}}-
\frac{3}{4 (\alpha \bar{c})^2}+ ...  \right), 
$$
for large $\bar{c}$ and
$$\bar{v}(\bar{c}) =  
\frac{\sqrt{\pi} \ (\beta V_o)^2}{L}  \left( \alpha \bar{c} -
 \sqrt{\pi} ( \alpha \bar{c})^2 + 3  ( \alpha \bar{c})^3 + ....\right),
$$
low  $\bar{c}$; with a maximum $\bar{v}_{max}=0.4346  (\beta V_o)^2/L$
at $c_{max}=0.82132/\alpha$.
 
  The results in the inset of Figure 1(b), shown that for 
$\beta V_o = 1$ the linear
response approximation for $\bar{v}(\bar{c})$ is very similar to the
exact result. In general, we may expect such good agreement for
gaussian barriers up to barrier height 
$\beta V_o \approx (\alpha/L)^{-1/2}$, beyond 
this limit  the linear response for low  drifting rates would
predict  unphysical values of $\bar{v}(\bar{c})> \bar{c}$, 
beyond the complete drift limit; which would indicate the 
breakdown of such linear response analysis since would predict
negative densities at the potential barrier.
However, for large drifting rates, when $\bar{v} \sim 1/\bar{c}$, the 
predictions of the linear response approximation are
quite good even for very large values of $\beta V_o$,
as shown in the inset of Figure 1(a).
The optimum value  $\bar{c}=\bar{c}_{max}$ to get the maximum $\bar{v}$ 
would be independent of $\beta V_o$ in the linear response
approach, while the exact result gives $\bar{c}_{max}$ growing
roughly linearly with $\beta V_o$, as the indicated by the 
crossover between the regimes $\bar{v} \approx \bar{c}$ and
$\bar{v} \sim (\beta V_o)^2/(L \alpha \bar{c})$, for low and high 
values of $\bar{c}$, respectively. However, the maximum
value of $\bar{v}$ from the exact result remains rather close
to the predicted by the linear response,  even at barriers as
high as $\beta V_o=20$, despite the different predictions
for the value $\bar{c}_{max}$ at which that  $\bar{v}_{max}$
is achieved. This behaviour is clearly shown in 
Figure 2, with logarithmic scales for both $\beta V_o$,
$\bar{v}_{max}$, $\bar{c}_{max}$ and the slope of 
$\bar{v}(\bar{c})$ at  $\bar{c}=0$, and it may be explained
as the results of cancellations of two different effects:
a tendency of the exact $\bar{v}(c)$ to grow faster with $\beta V_o$ than the
prediction of the linear response, for low barriers, and 
the tendency of $\bar{c}_{max}$ to grow linearly with $\beta V_o$
at high barriers, leading to $\bar{v}_{max} \sim \bar{c}_{max} 
\sim \beta V_o/\sqrt{\alpha L}$ for very large $\beta V_o$.
Similar results, with different analytic forms for $\bar{v}(\bar{c})$ 
may be obtained for the  parabolic barriers or other parametric forms of
the external potential.

  In Figure 3 we present the density profiles and the steady currents
for a smooth gaussian  barrier 
of width  $\alpha=0.707\sigma$, space period $L=30\sigma$, and two set of the
barrier heights  and the shifting rates. For $\beta V_o=1$ the results of 
the linear response theory gives  a reasonably accurate description of the 
exact density profiles for the slowest shifting rate, $\bar{c}=1\sigma^{-1}$ in
Fig. 3(a) and a very accurate results at $\bar{c}=10\sigma^{-1}$ in
Fig. 3(b). The shape of $\tilde{\rho}(\hat{x})$, 
relative to the potential barrier (shown by the dot-filled gaussians)
reflects the difference between 
the quasi-static behavior for $\bar{c}=1\sigma^{-1}$, and the strong effects of 
the kinetic effective potential $V_k(\hat{x})=V(\hat{x})+\bar{c} \hat{x}$
at $\bar{c}=10\sigma^{-1}$, which pushes the maximum of $\tilde{\rho}(\hat{x})$
toward the maximum of $V(\hat{x})$. The current, which may be represented 
by the same line with the appropriate changes in the vertical axis to
get (\ref{jota1}), has a positive contribution in front of the barrier
partially cancel by the negative contribution across the 
barrier.

For higher barriers , $\beta V_o=5$, and low drift rate  Figures 3(c), the
system becomes close to the complete drift regime ($\Delta
\rightarrow 0$), with a strong accumulation of particles in front of
the advancing barrier, which decays exponentially
$\tilde{\rho}(\hat{x}) \sim exp(- \bar{c} \hat{x})$, so that 
for long distances $L$ between the periodic barriers the density
at the receding side is extremely low compared with the 
density at the advancing side.  
The linear response results are quite out of 
the mark, predicting unphysical negative density at the barrier.
At higher values of $\bar{c}$   Figures 3(d) the
accumulation of particles at the barrier front is so large that the 
barrier becomes ineffective to keep the complete drift regime, the
local chemical potential, $\mu(\tilde{\rho})=\beta^{-1} log(\tilde{\rho})$
becomes comparable to $V_o$ and the particles jump over the barrier.
The final effect is that the counter-drift $\Delta$ grows and the 
total effective potential in (\ref{DDFeq1}) becomes much weaker
that the external potential $V(\hat{x})$. In this way, the 
current decreases and it becomes closer to the linear
response result, with the limit of no density perturbation
or steady current for very rapid displacements of the barrier
( i.e.  $\bar{c}\rightarrow \infty$), for any smooth shape of the
barrier.

\section{ Steady state for hard-rod particles}

 In the solution of the DDF equations for the ideal gas case
the mean density of the system is irrelevant for the mean velocity
$\bar{v}$, as both the density $\rho(\hat{x})$ and the current
$j(\hat{x})$ have the common factor $A$ to include the constrain 
of fixed number of particles. In this section we explore 
the effects of molecular packing and correlations on the stationary
currents created by the shifting external potentials $V(x- c \ t)=V(\hat{x})$.
For particles moving along a linear channel, the choice of hard-rods
(with length $\sigma$)
to model the interactions allows the use of the exact DF for the
equilibrium free energy \cite{Percus}. The functional  
 derivative of the equilibrium free energy in (\ref{DDF}) is
\begin{equation}
\frac{\delta \beta{\cal F}[\rho]}{\delta \rho(x)}= log\left( \frac{\rho(x)}
{1- \eta(x)}\right) + \int_{x-\sigma}^x dx'  \frac{\rho(x')}
{1- \eta(x')},
\label{dfdr}
\end{equation}
with the local packing fraction $ \eta(x)$ defined as
\begin{equation}
\eta(x) = \int_x^{x+\sigma} dx'  \rho(x').
\label{etax}
\end{equation}

The equation(\ref{DDFeq1}) for the stationary density becomes now
an integral equation,
to be solved with periodic boundary conditions and the
the self-consistency requirement (\ref{Delta}). The total
number of particles in a period $N= L \langle \rho(\hat{x}) \rangle $,
should be controlled by the chemical potential $\mu$, with non-trivial
dependence of $\rho(\hat{x})$, $\Delta$ and $j(\hat{x})$.
Alternatively, we may solve dynamic equation (\ref{DDF}) for $\rho(x,t)$, 
from a reasonable guess for the initial density distribution at $t=0$,
and integrate in time until the stationary limit $\rho(x,t)=\rho(x - c \ t)$
is achieved. In that case, the number of particles is set with the 
initial guess (since equation (\ref{DDF}) keeps constant $N$).
We have checked that both procedures lead to the same stationary results, 
and the relative computational efficiency would depend on the characteristics
of the external potential, and the choice of the initial profile.

  For weak perturbations (weak external potentials or
very large drifting rates) the linear response analysis (\ref{linear}), has to 
be applied with the hard-rods structure factor,
$$
S(x,\rho)= \left[ 1 + 2 \frac{\eta}{1-\eta} 
\frac{sin(q \sigma)}{q \sigma} + 
2 \left(\frac{\eta}{1-\eta} \right)^2
\frac{1- cos(q \sigma)}{(q \sigma)^2} \right]^{-1},
$$
with the packing fraction  $\eta =\sigma \rho_o$. Near 
the complete packing limit $\eta \rightarrow 1$,
the effect of a shifting  oscillatory external potential
$V(x,t)=V_q  cos(q (x - c t))$, would depend 
strongly on the wave vector $q \sigma$; as shown in Figure 4(a) 
for constant value of $V_q$ (i.e. constant amplitude for the 
oscillation of the external potential), the optimal choice of $q$
to get maximum mean velocity $\bar{v}$, at  low values of $c$,
corresponds to $q$ at the 
first peak of the structure factor, just below $q \sigma= 2 \pi$,
with a large enhancement over the velocity obtained in an ideal
gas with the same external potential. For larger values of $c$
the maximum $\bar{v}$ is obtained for $q$ in the second or in 
the successive peaks of $S(q,\rho_o)$.  The results in Figure 4(b)
for constant $q V_q$ (i.e. constant amplitude of the oscillating force)
indicate that the optimum choice in that case is always at the first 
peak of the structure factor, but the enhancement over the ideal gas
is strongly reduced as $c$ increases.

  For well separated periodic barriers, like gaussians of
width $\alpha$ much larger than their period $L$, the main effects
of the hard-rod interactions are to decrease the compressibility 
and to form layered structures on both sides of the 
barrier, as usual in the equilibrium density distributions of a
hard-core fluid near a wall.  In figure 5 we present the results
for a dense system, with mean density $N/L=0.8 \sigma^{-1}$,
under the action of high gaussian barriers ($\beta V_o=10.6$ and
width $\alpha=0.707 \sigma$, separated by a distance $L=30 \sigma$.
At low shifting rate ($\bar{c}=0.2 \sigma^{-1}$ in Fig. 5(a) )
the system is close to the complete drift limit, with very low 
density at the potential barrier. On both sides of the barrier
here are strong oscillatory structures, around a nearly constant 
density. The amplitude of the oscillations is larger at the advancing
side than at the receding side of the potential, but this may be
understood in terms of the equilibrium density profiles for the 
effective kinetic potential 
$V_k(\hat{x})=V(\hat{x})+ \beta^{-1} \bar{c} \hat{x}$,
which makes sharper the effective wall at one side and smoother at
the opposite side.

  Increasing $\bar{c}$, in Figures 5(b-c), produces an apparently weak 
effect on the density at the outside the barrier. For comparison 
we include the ideal gas results, for the same mean density and barrier
$\beta V_o=5$, chosen to be  equivalent to the   effective barrier in the hard rod,
with the recipe describe belong ,which have high peak densities at 
the advancing front with maximum values of $\tilde{\rho}_{max}=3.3  \sigma^{-1}$ for $\bar{c}=1 \sigma^{-1}$
and$\tilde{\rho}_{max}= 7.5 \sigma^{-1}$ for $\bar{c}=1.7 \sigma^{-1}$,
with exponential decays toward a much lower on the receding edge of the 
next barrier.
The low compressibility of the dense hard-rod fluid strongly reduces such
accumulation of particles in front of the barrier and their sensibility
to the value of $\bar{c}$; the main change between the results for
hard-rods in  Figures 5(a) and 5(b) is in the
increase of the minimum density , near the barrier maximum,
which indicates that the local chemical potential may get easily over 
$V_o$ with very weak changes in the local density, as the compressibility 
of the system $(\beta \rho)^{-1} d\rho/d\mu=(1-\eta)^2$ produces strong
changes in $\mu$ with small changes in $\eta$, and the particles 
start jumping over the barrier without building up the high local densities
of the ideal gas. Only for the largest value of the shifting rate,
$\bar{c}=10\sigma^{-1}$ in Fig. 5(d),  we start to see a clear accumulation of
particles at the advancing front, but still being a small fraction 
of the total number of particles in the system. The oscillatory
structures become damped, with respect to their shapes for lower
$\bar{c}$ and this effect is stronger on the advancing than on the
receding front, as a signature that the effect of term proportional to
$\Delta$ in the effective potential of the Euler-Lagrange equation
(\ref{DDFeq1}) is getting over linear kinetic contribution 
$\bar{c} \hat{x}$.
The linear response results, presented as the dotted lines in Figures 5,
are quite  different from the full DDF calculations, as could
be expected for such a high barrier, except for the largest value of
$\bar{c}=10\sigma^{-1}$. However, even in that case the linear response
fails to reproduce the layering structure.

  Finally, we present in Figure 6, the mean velocities produced 
by gaussian barriers, as functions of the shifting rate, in system 
of different mean density. The results for the ideal gas with
a barrier height of $\beta V_o=5$ are compared with those for
mean densities of $\rho_o=N/L=0.2$, $0.4$, $0.6$ and $0.8\sigma^{-1}$,
with barrier heights adjusted to give the same effective barrier,
$V_o-\Delta \mu(\rho_o)$, in terms of the excess chemical potential
over the ideal gas, at each mean density. This semi-empirical recipe
is effective to give similar values of $\bar{v}$ for different
$\rho_o$, while keeping the same absolute barrier height produces 
a rapid decrease of $\bar{v}$ with the mean density.
In all the cases the qualitative shape of $\bar{v}(\bar{c})$ is similar
to that of the ideal gas, with a linear growth, $\bar{v} \approx \bar{c}$
in the quasi-complete drift regime, followed by a maximum mean velocity
and a decrease $\bar{v} \sim 1/ \bar{c}$ at large shifting rates.
The particular values of $\bar{v}_{max}$ are tied to our empirical
choice of the values of  $\beta V_o$ at each mean density, but the
optimal rate $\bar{c}_{max}$ at which it is achieved, and the generic
shape of the curve are roughly independent of the barrier height.
In this respect, the main effect of the hard-core interactions seems
to be the broadening of the maxima which becomes much less sensitive
to the value of $\bar{c}$ than for ideal gas systems.
For very large shifting rates the linear response analysis becomes 
valid again, as for ideal gas systems, and it becomes independent 
of $S(q,\rho)$ and hence equal to that for the ideal gas.
This may be interpreted, in physical terms, as the result of a very 
rapid series of of pushing and pulling forces, acting on the particles
as the potential barriers go over them, with a time period much shorter
than the typical collision time between particles; so that the later
become irrelevant.

\section{ Conclusions}

 We have applied the dynamic density functional (DDF)
theory to study the relaxative dynamic of particles moving along  
one-dimensional channels, under the action of external potential
shifting at constant rate. Our results could be of interest for 
systems of colloidal particles, with typical sizes of $\sigma \sim 10^{-7}m$,
following brownian motion on water at room temperature. In that case 
the relevant values of the shifting rates of the external potential
would be around $10^{-4}m/s$, which could be achieved by the   
the dielectric force created on the colloidal particles by a moving 
laser beam or the force created on surface charged colloidal particles 
by a moving electric potential. Although strongly simplified with 
respect to the complexities of real micro-fluidic devices, we may
hope that the model and its theoretical treatment contains the essential
features of real systems. In principle the model could be studied 
directly from the stochastic equation for the brownian motion of
the particles, without relaying on the equilibrium density functional
formalism and its approximate application to the description of 
systems out of equilibrium. For the ideal gas case the results should
be identical, but the DDF theory allows a direct extension to interacting 
systems, which we have considered here as pure repulsive interactions
treating the colloidal particles in a narrow channel as hard rods
particles in one dimension.

  Under the effects of a periodic external potential shifted at 
constant rate $c$, the systems would achieve a steady state, with 
density distributions and local currents following the shift of 
the external potential. These are the only problems analyzed in this
paper, leaving for future works the analysis of the transient evolutions
to achieve such steady state from arbitrary initial conditions. 
The steady states are particularly appropriate to be studied within the 
DDF, since the structure of the relevant equations becomes similar to that
of the usual density functional formalism for the equilibrium properties.
Thus, the generic  equation (\ref{DDFeq1}) is equivalent to an
Euler-Lagrange equation for the equilibrium density distribution
but including two extra terms as external potentials. The first one,
which we have called the {\it kinetic} contribution, is just a
linear potential, proportional to the shifting rate $c$ which breaks
the symmetry between the advancing and the receding front of
a shifting potential barrier. The second contribution to the effective 
external potential is a {\it self-consistent} term which depends on
the integral of the inverse density distribution over the system,
and it is proportional to the second integration constant, $\Delta$, 
which (together with the chemical potential $\mu$) is required
to get the general solution of the integro-differential Euler-Lagrange
equation for the problem. For external potentials made of periodic 
barriers, we have given an intuitive interpretation to $\Delta$ as
the {\it counter-drift} current, associated to those particles 
which jump over the barriers and so reduce the mean current in the
system, with respect to the {\it complete-drift} regime in which
the full density is shifted at the same rate as the barriers, and the 
density distribution may be obtained as the equilibrium profiles for 
the fluid in presence of the external plus kinetic potentials.
This regime would appear, and hence the integration constant $\Delta$
would vanish, whenever the potential barriers are large enough, compared
with the kinetic effective potential.  In the opposite limit, 
for very weak shifting external potentials, we have extended to the DDF the
usual linear response analysis of the equilibrium density functional
formalism. The range of application for such analysis is in fact
enlarged with respect to the equilibrium case, since strong external
potentials moving at very large rate would produce much weaker effect
than the same external potentials at rest. In that limit, 
opposite to the complete-drift, the self-consistent contribution
to the effective potentials cancels out most of the linear {\it kinetic}
term and a good part of the external potential, so that the density
distribution is much smoother as a function of the position that 
the static equilibrium density profile. Only for discontinuous
external potentials, like square barriers, there would be a
remnant effect of the shifting external potential in the limit
of $c \rightarrow \infty$, and it should be considered as an artifact
of the infinite forces acting on the particles at the jumps of $V(x)$.  

  Although we have used periodic potential barriers in our
analytical and numerical solution of DDF problems, we have
also considered the case of single potential barriers as
the $L \rightarrow \infty$ limit of the periodic potentials.
In that case, the integration constant $\Delta$ goes to
the shifting rate $c$ times the bulk density of the fluid, $\rho_o$
far away from the moving external potential. This is the requirement 
to get null current away from the external perturbation as the only
possible solution for the stationary density distribution in an infinite
one-dimensional system with a shifting potential acting on a finite
region. Any solution of the Euler-Lagrange equation with $\Delta \neq c \rho_o$
would imply the accumulation of particles at the advancing front of
the potential barrier and the depletion of the density behind it,
so that the system would not be at any  stationary state. In any 
realistic problem, like possible devices to control the flow of 
colloidal particles along channels, the finite length of the 
channels would probably produce typical density distributions and currents
closer to those analyzed here for periodic systems than to those
predicted for infinite long channels.

  For the periodic systems studied here, we have found that the
mean velocity of the particles may be maximized by the 
appropriate choice of the shifting rate. The hard core 
interactions among the particles reduces the optimal
value of $\bar{v}_{max}$, but makes this mean velocity 
a broader function of the shifting rate. At high packing fractions,
$\bar{v}_{max}$ has a strong and  non-monotonic dependence on the period 
of the external potential, as a results of the correlations between 
particles. Such effect could be used to design more effective 
micro-pumps or turn-pikes, in which the flow of colloidal particles
would be controlled by the oscillatory external potential.
The particular models studied here, ideal gas and hard-rods,
under the action of gaussian potential barriers or harmonic 
periodic potentials, have been chosen to explore the main
regimes and generic properties of stationary flows in one-dimension.
There are obviously many open questions to answer and different
models to be explored. The application of the density functional
techniques to these systems is still at its beginnings, and it would
take time to ascertain its full capabilities. However, the generic
theoretical requirements: the absence of hydrodynamic effects and
the similarity between the particle correlations in the dynamic
and in the equivalent equilibrium systems, seem likely to be reasonable
assumptions in systems of colloidal particles moving at rates of the 
order of microns per second, under the effects of controlled external
potentials. The application of the DDF techniques to systems in higher
dimensions and to non-stationary problems would hopefully benefit from 
what we have learned with our simpler models and problems. The design 
of practical micro-fluiditic devices could also be supported by the
generic principles developed with these simple models, and the 
experimental realization of these systems appears to be within the
range of the present technologies. 

\section{Acknowledgments}
We acknowledge financial support
by the Direcci{\'o}n General de Investigaci\'on,
MCyT, under grant number BFM2001-1679-C03-02 and 
by a FPU-grant number AP2001-0074  from the MECD of Spain.

\newpage

\begin{figure}
\caption{Mean velocity of the particles as a function of the barrier
drift rate, for square barrier (solid line) of width $a=2\sigma$;
parabolic barrier (dotted line) of width $w=5.5\sigma$ and gaussian barrier (dashed) 
of width $\alpha =1\sigma$. The  barrier heights are $\beta V_o=7$ in figure (a)
and  $\beta V_o=1$ in figure (b).The insets compare,  for
the gaussian barriers, the exact results (solid lines) with the linear 
response theory (dashed lines).
}
\label{fig:1}
\end{figure}    
\begin{figure}
\caption{Maximum  mean velocity value $\bar{v}_{max}$, 
drift rate, $\bar{c}_{max}$,
at with this maximum mean velocity value is achieved, and the 
derivative of  mean velocity at zero  drift rate, $\bar{v}'(\sigma\bar{c}=0)$,
for a gaussian barrier as functions of the barrier height.
The solid lines give the exact values and the dashed lines
give  linear response approximation, which are similar to the exact
for low barriers.
}
\label{fig:2}
\end{figure}
\begin{figure}
\caption{Exact density distributions for an  ideal gas system  
(solid lines  referred to the left axis) and  currents
(the same lines but referred to the right axis), for gaussian
barriers of width  $\alpha=0.707\sigma$ and heights $\beta V_o=1 $(a) and (b) and  $\beta V_o=7 $(c) and (d)   .
The drift rates are  $\bar{c}=0.5\sigma^{-1}$ for (a) and (c), and  $\bar{c}=10\sigma^{-1}$
for (b) and (d). The dot-filled lines represent the external potential barrier
(not at scale) just to show the relative position of the density structures
to the barrier. The dashed lines (referred to the left axis) give
the density distributions predicted by the linear response theory.
}
\label{fig:3}
\end{figure}
\begin{figure}
\caption{Mean velocity of the particles under a shifting oscillatory
external potential as a function of its wave-number q, 
for constant value of the barrier amplitude, $\beta V_q=1$, in  (a),
and constant force amplitude   $q \sigma \beta V_q=1$ in (b).
The full line gives the result for hard-rods at $\rho_o=0.8\sigma^{-1}$
with $\bar{c}=2 \sigma^{-1}$, and the dashed lines the same for  
$\bar{c}=10 \sigma^{-1} $.
The dotted lines gives the ideal gas (i.e. $\rho_o \sigma =0$)
under the same drifting potentials for  $\bar{c}=2\sigma^{-1}$ and 
the dash-dotted line for $\bar{c}=10\sigma^{-1} $. All the results
are calculated with the linear response theory.
}
\label{fig:4}
\end{figure}
\begin{figure}
\caption{Density distributions (solid lines referred to the left axis) 
and  currents (the same lines but referred to the right axis),
for  hard rods of mean density $\rho_o=0.8\sigma^{-1} $ under the 
influence of an external gaussian potential of height $\beta V_o=10.6$ and width  $\alpha=0.707\sigma$.
The density distribution predicted by the linear response theory
(dotted lines) and the ideal gas for  an external gaussian potential of height $\beta V_o=5$ and  width  $\alpha=0.707\sigma$ (dashed lines) are
also  presented for comparison.
The drift rates are
$\bar{c}=0.2\sigma^{-1}$ in (a), $\bar{c}=1.0\sigma^{-1}$ in (b), $\bar{c}=1.7\sigma^{-1}$ in (c), and
$\bar{c}=10\sigma^{-1}$ in (d).
}
\label{fig:5}
\end{figure}
\begin{figure}
\caption{Mean velocity of hard-rod particles, as function of the
drift rate, $\bar{c}$,  produced by a gaussian barrier of high 
$\beta V_o=5$  in  an ideal gas system, and a gaussian barriers of height
$\beta V_o=5.47$, $6.18$, $7.41$, and $10.6$ for systems with mean 
density $\rho_o \sigma=0.2$, $0.4$, $0.6$, $0.8$ respectively.The width
of all gaussian is $\alpha=0.707\sigma$.
}
\label{fig:6}
\end{figure}

\end{document}